\documentclass[%
 reprint,
 amsmath,amssymb,
 aps,
]{revtex4-2}

\usepackage{graphicx}
\usepackage{dcolumn}
\usepackage{bm}
\usepackage{hyperref}
\usepackage[mathlines]{lineno}

\begin{document}

\preprint{APS/123-QED}

\title{The role of compressional dynamics in setting the scale-dependent rheology of granular flows: Application to the emergence of thin layer stability}

\author{Christopher Harper}
 \email{charper4@uoregon.edu}
 
\author{Josef Dufek}%
\thanks{These authors contributed equally to this work}
\affiliation{%
Department of Earth Science, University of Oregon
}%

\author{Eric C.P. Breard$^\dagger$}
\altaffiliation[Also at ]{Earth Science Department, UO}
\affiliation{%
 School of Geosciences, University of Edinburg
}%

\date{\today}

\begin{abstract}
One great challenge of modeling granular systems lies in capturing the rheologic dependencies on scale.  For example, there are marked differences between quasi-static, intermediate, and rapid flow regimes. In this study, we demonstrate that assumptions for infinite stiffness of rigid particles, an assumption upon which the state-of the-art ($\mu(I)$-rheology) modeling approaches are constructed, must be relaxed in  order to recover the physical mechanisms behind many scale-dependent and non-local rheological effects.  Any  relaxation of the infinite stiffness assumption allows for particles to compress in series, whereby the number of simultaneously compressed particles controls the extent to which end-member particles experience a modified coefficient of effective friction, analogous to reduced stiffness for springs in series.  To demonstrate the importance of such a mechanism in setting the dynamics for dense rigid granular systems, we show that modifying simple models to include the kinematics introduced by compression in series captures the emergence of thin layer stability, a widely observed yet incompletely explained non-local granular phenomenon. We also discuss, in general, how  knowledge of the contact network and softness provides a potential physical basis for the diffusion of granular temperature.
\end{abstract}

\maketitle


\section{\label{sec:level1}Introduction}

Granular processes are as ubiquitous as they are complex, present in the bases of fast flowing pyroclastic density currents, and the slow shifting stability (creeping) of slopes. The governing dynamics of these two end-member systems have disparate characterizations; the former better described by kinetic theory and binary particle collision, while the latter is dominated by enduring grain contacts, friction, and slow deformation.  Despite  different characterizations,  both dynamics can manifest simultaneously in even the simplest granular systems, as demonstrated in Rocha's (2019) experiments, in which flows of mostly mono-disperse inclined granular systems exhibit a fast flowing center guided by slowly shifting to nearly stationary edges (Ref.~[\onlinecite{rocha_self-channelisation_2019}]). One of the great challenges of describing granular systems lies in finding the physical mechanisms capable of producing such complex rheologies, which, while essential for accurate modeling,  are poorly understood. Underlying much of these difficulties is a scale dependent rheology which makes direct analogies to solid, liquid, and gas phases of matter difficult (Ref.~[\onlinecite{jaeger1996granular}]) . This complexity is in part because the grains form a  fabric of contacts which allows for grains, not directly in contact with one another, to significantly affect each other (Ref.~[\onlinecite{Papadopoulos_2018}]).  This lengthscale at which non-local particles can affect one another is referred to as the mesoscale, and the effects which manifest from particle contacts at the mesoscale are typically referred to as non-local effects. 

Despite these complexities, an insightful argument by da Cruz (2005)~[\onlinecite{DaCruz2005}] showed that evolution of stresses in systems of simply sheared infinitely rigid grains are well characterised by a single dimensionless parameter, the inertial number``$I$". Surprisingly this global description for simple shear could be used to describe the evolution of stresses locally in inhomogeneous systems giving rise to a powerful rheological frame work known as the local $\mu(I)$-rheology (Ref.~[\onlinecite{Forterre2008}]).  This framework has been successfully expanded to capture a wide range of dense granular systems (Refs.~[\onlinecite{boyer_unifying_2011, amarsid_viscoinertial_2017}]), however these approaches universally struggle with the scale-dependent dynamics which emerge in granular domains with low inertial numbers: such as transition to flow, jamming, creep, and many other non-local effects. 

Phenomenological descriptions, such as the kinetic elasto-plastic (KEP) theorem (Ref.~[\onlinecite{KEP}]) or granular fluidity (Ref.~[\onlinecite{HENANN2014145}]), have been proposed to modify the $\mu(I)$-rheologic framework and capture quasi-static granular behavior. However, a kinematic state variable which can describe the physical underpinnings of these phenomenology remains elusive. In this work we argue that assumptions for infinite stiffness of rigid particles, an assumption upon which the $\mu(I)$-rheology is predicated, must be relaxed in  order to recover such physical mechanisms. We claim that by allowing for any epsilon of softness, however slight, and therefore deformation,  our ``rigid" particles will compress in series introducing an effective stiffness which modifies local coefficients of effective friction for end-members. This effective stiffness is analogous to the effect seen for $n$ springs in series captured in equation \ref{eq:effective_k}.

\begin{equation}
    k'=\frac{k}{n}
    \label{eq:effective_k}
\end{equation}

Notice that the effect remains significant in the limit and only vanishes at infinite stiffness ($k=\infty$).  Thus any physical regimes where we this effect is expected to be significant are likely to be difficult for the $\mu(I)$-rheology to fully capture. To demonstrate the importance of compression in series, also referred to as compression chains or chains of compression, we show that capturing such an effect in a simple kinematic model of inclined granular flow explains the emergence of a widely observed yet still incompletely explained granular phenomenon - thin layer stability. Thin layer stability simply states that the critical stresses for onset and arrest of flow depend on a non-dimensionalized system height (hereafter referred to as height), and is a phenomenon which exhibits both scale dependence and non-locality.  Understanding thin layer stability is essential to being able to explain many other perplexing granular phenomenon;  it manifests in the clogging of silos (Ref.~[\onlinecite{annurev_karmin}]) and in levee formation, which can lead to self channelization and increased run out observed in many large-scale and complex geophysical flows (e.g. lava flows, pyroclastic density currents, debris flows etc) (Refs.~[\onlinecite{mangeney2007numerical, sparks1976classification, kokelaar2014fine, jessop2012lidar, johnson2012grain}]). 

We capture the effects of deformation on the kinematics of inclined granular flow within a modified "grain on a sandpile model" (Ref.~[\onlinecite{quartier_dynamics_2017}]) coupled with our own deformable collision model. The results are compared to Perrin's experimental results for inclined planar granular flow of frictionless particles (Ref.~[\onlinecite{Perrin2021}]), and show an excellent recovery of the observed thin layer stability. We conclude with a discussion for how this phenomenon manifests more generally, and when coupled with knowledge of contact networks could provide a physical basis for the phenomenological notions of granular fluidity and the transport of granular temperature.

\section{\label{sec:level1}Background}
\subsection{\label{sec:TLS}Thin Layer Stability}
\begin{figure}[h]
\includegraphics[width=.45\textwidth]{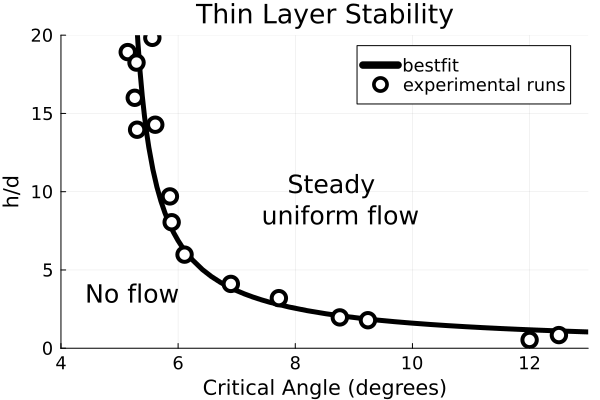}
\caption{\label{fig:perrin_TLS} Perrin's observations of thin layer stability manifesting in systems of frictionless grains Ref.~[\onlinecite{Perrin2021}]). The open circles represent results measured from experimental runs and the black is best fit estimate. Critical angle denotes an angel of inclination for the system and $h/d$ is a dimensionless height.}
\end{figure}

Thin layer stability, an incompletely explained non-local effect, is most simply exhibited in inclined planar flow, where it manifests as a dependence between the angle of repose and the non-dimensionalized system height (Ref.~[\onlinecite{pouliquen_1999}]). Thinner systems of grains have a higher angle of repose, with this critical angle decreasing asymptotically as system height increases. This phenomenon has a long history of experimental observations in a variety of granular systems, ranging from glass beads to walnut shells (Ref.~[\onlinecite{pouliquen_1996}]).  This breadth of systems implies a general importance to mesoscale dynamics and significance to non-local effects in general, while the simplicity of systems in which it can manifest emphasize a link to fundamental granular behavior; thin layer stability has been observed in inclined planar flows of mostly mono-disperse frictionless grains (Figure \ref{fig:perrin_TLS}). 

Without a material coefficient of friction, Perrin's granular systems can only resist the driving force of gravity through particle geometry and contact structures, and in such a system the coefficient of  effective friction reduces to the geometric friction. This removes confounding effects, and places a heavy burden on any proposed structural mechanism to be the primary driving force behind the observed scaling in figure \ref{fig:perrin_TLS}.  This simplicity offers an ideal and exacting framework for evaluating our proposed physical mechanism of compression chains. The experimental findings of Perrin et al. (2021)(Ref.~[\onlinecite{Perrin2021}]) will serve as the benchmark for assessing our model which is built off of a simple theoretical framework pioneered by Quartier(Ref.~[\onlinecite{quartier_dynamics_2017}]).

\subsection{\label{sec:quartier}The Quartier Model: a simple theoretical framework}
\begin{figure}[h]
    \includegraphics[width=.45\textwidth]{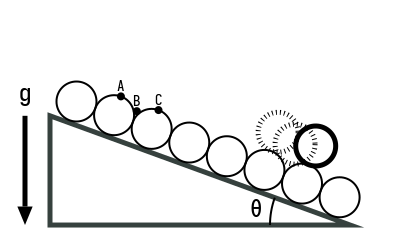}
    \caption{A schematic of the Quartier system; a single gravity driven frictionless grain slides down a substrate of identical but fixed grains at an incline of $\theta$ .}
    \label{fig:quartier_model}
\end{figure}

The Quartier model consists of one mobile grain and a substrate of fixed grains on an inclined plane, as shown in figure \ref{fig:quartier_model}.  A complete analysis of this system was done in Quartier's  publication on the  ``Dynamics of a grain on a sandpile". Despite its simplistic nature the analysis clearly reveals promising roots for many complex granular phenomenon, such as the  ability for a substrate of grains to form  periodic potential traps, how these traps, in turn, bifurcate the system into a static and dynamic steady state, and how simple geometric constraints can give rise to hysteresis (we will highlight this behavior later)(Ref.~[\onlinecite{quartier_dynamics_2017}]).  The analysis performed here involves frictionless particles to be consistent with the experiments of Perrin et al. (2021)(Ref.~[\onlinecite{Perrin2021}]), and can take a more simplified approach than the original work. For clarity, we refer to the mobile particle as a grain, and the fixed substrate as particles. 

In this frictionless analysis, the grain does not roll, but slides over the surface of the glued particles as it is driven by gravity. It is imbued with an initial kinetic energy $E_0$, such that it never becomes airborne, and in this way the dynamics of the system are determined by the interactions at points A,B, and C.

At point A the grain is sliding down the crest of one particle and into the gravity well at point B. The grain will then crest the next particle at point C.  Whether the grain successfully crests the next particle depends on a balance between the kinetic energy gained from cresting the previous particle, the cost of escaping the energy well ($\mathcal{C}$), and the energy lost in the shock experienced at point B ($\mathcal{E}$).  The first two are solely dependent on system geometry and summarized in table \ref{tab:key_values}, the ability for the particle to remain mobile is determined by the interaction at point B, which bears further scrutiny. 

\begin{table}[h]
\caption{\label{tab:key_values}%
Key variables in the quartier model
}
\begin{ruledtabular}
\begin{tabular}{lcdr}
\multicolumn{1}{c}{\textrm{Key system parameters}}&
\textrm{expression}\\
\colrule
 initial kinetic energy & $E_0$ \\
gravity well & $\mathcal{C} = mgr\left(1- cos\,(\frac{\pi}{6} -\theta)\right)$ \\
kinetic energy gained per crest  & $\Delta E = mgr\cdot sin\,\theta$ \\
 energy after first shock & $E_1 = (E_0 + \Delta E)\mathcal{E}$ \\
\end{tabular}
\end{ruledtabular}
\end{table}

\begin{figure}[h]
  \centering
  \begin{minipage}[b]{0.22\textwidth}
    \centering
    \includegraphics[width=\textwidth]{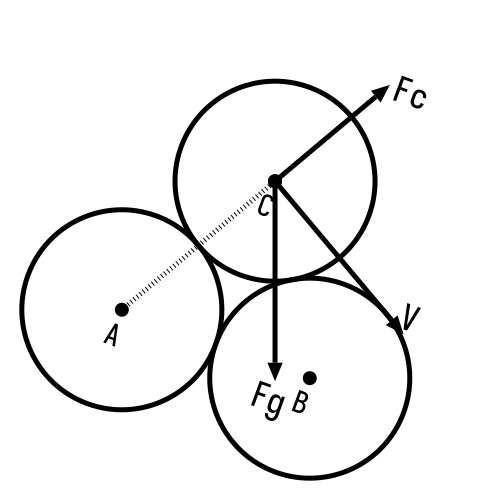}
    \textbf{(a)} Pre-shock
  \end{minipage}
  \hspace{0.015\textwidth}
  \begin{minipage}[b]{0.22\textwidth}
    \centering
    \includegraphics[width=\textwidth]{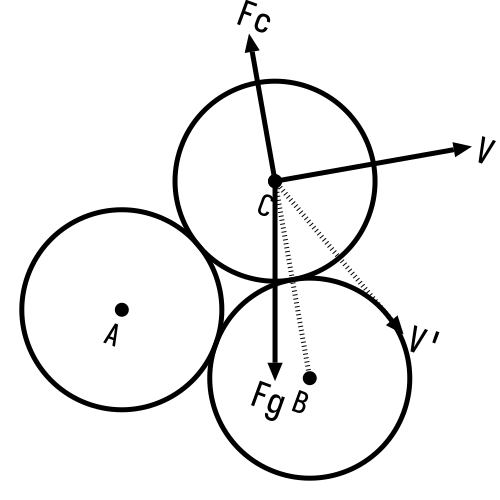}
    \textbf{(b)} Post-shock
  \end{minipage}
    \caption{(a) The force diagrams for the mobile grain instantaneously before and (b) after shock. The dashed illustrates the shifting particle contact. V' represents the velocity preshock, and is shown in (b) to emphasize the acceleration forced by system geometry.}
  \label{fig:force_diagrams}
\end{figure}

Figure \ref{fig:force_diagrams} shows the force diagram for the mobile grain right before and after the shock experienced at the bottom. During the shock, the velocity accelerates from being perpendicular to particle contact AC (panel(a)) to perpendicular to particle contact BC (panel (b)), and some energy is dissipated and captured by the term $\mathcal{E}$ (kinetic energy after the shock/kinetic energy before the shock). By assumption of constant contact, the contact force ($F_c$) can never be larger in magnitude than the normal component of the force of gravity, otherwise the grain would become airborne.

While this might seem restrictive, it allows for a powerful analysis of the system by expressing the energy at each successive shock in a recursive manner (last line of table \ref{tab:key_values}). If the energy after the first shock is sufficient to escape another well ($\mathcal{C} \le E_1$), then  we can describe the energy after the second shock as $E_2 = (E_1 + \Delta E)\mathcal{E}$ . Expanding and simplifying for successive terms reveals the following pattern:

\begin{equation}
    E_n = E_0\cdot\mathcal{E}^{n}+\Delta E\cdot\mathcal{E}\underbrace{\left(\mathcal{E}^{n-1}+\mathcal{E}^{n-1}+...+\mathcal{E}^2+\mathcal{E} +1\right)}_{S_n}
\end{equation}
The part labeled $S_n$ can be simplified using the following series identity.
\begin{equation}
    S_n = 1+x+x^2+...+x^n = \frac{1-x^{n+1}}{1-x}
\end{equation}
Resulting in a useful closed form expression for the energy after the $n^{th}$ shock:
\begin{equation}
    E_n = E_1\cdot\mathcal{E}^{n-1} +\Delta E\cdot\mathcal{E}\left(\frac{1-\mathcal{E}^{n-1}}{1-\mathcal{E}}\right)
\end{equation}
If the energy gained from cresting a particle is greater than the energy dissipated through the shock ($\mathcal{E} > 1$),  $E_n$ will go to infinity, and this grain will accelerate indefinitely. However, if the energy dissipated from the shock is greater than the energy gained, then the system will asymptotically approach a constant energy state:
\begin{equation}
    \lim_{n\to\infty} E_n = \Delta E \left(\frac{\mathcal{E}}{1-\mathcal{E}}\right)
    \label{eq:energy_sequence}
\end{equation}

We can set this limiting energy equal to the expression for the cost of the energy well to find a critical point of our system, the angle at which our grain can first enter into steady state motion. This minimum angle is called the critical angle:

\begin{eqnarray}
    \Delta E \left(\frac{\mathcal{E}}{1-\mathcal{E}}\right) \geq mgr\left(1- cos\,(\frac{\pi}{6} -\theta)\right) \rightarrow \nonumber \\
     \frac{\mathcal{E}}{1-\mathcal{E}}\geq \frac{2sin^2(\frac{\pi}{12}-\frac{\theta}{4})}{sin(\theta)}
    \label{eq:steady_states}
\end{eqnarray}

From equation \ref{eq:steady_states} we see that the critical angle is only dependent on the angle of inclination and the dissipative term, $\mathcal{E}$, which represents the ratio of kinetic energy after and before each shock(equation \ref{eq:shock}).  

\begin{equation}
    \mathcal{E}=|\frac{v_i'-v_i}{v_i'}|^2
    \label{eq:shock}
\end{equation}

The velocity instantaneously before and after the $i^{th}$ shock is represented as $v_i'$ and $v_i$ respectively. Using the geometric constraints of our system, and introducing a new material property, the normal restitution coefficient ($e_n$), we can solve for the velocity post-shock (equation \ref{eq:pre_post_shock}).

\begin{equation}
v_i = \sqrt{\left(\underbrace{v_i'\cdot\frac{1}{2}}_{\hat{i}}\right)^2+\left(e_n\left(\underbrace{v_i'\cdot\frac{\sqrt{3}}{2}}_{\hat{j}}\right)\right)^2}
\label{eq:pre_post_shock}
\end{equation}

This can be interpreted as the original velocity multiplied with the new component vector post shock, with $\hat{i}$ being the direction tangent to new particle contact, and $\hat{j}$ the direction perpendicular . In this form we can clearly see the role of the normal restitution  ($e_n$),  which represents the percentage of velocity normal to particle contact conserved after collision. The factors of $\frac{1}{2}$ and $\frac{\sqrt{3}}{2}$ stem from the equilateral triangle created from the geometry of mono-disperse particles in a configuration shown in fig \ref{fig:force_diagrams}.  By substituting this expression into $v_i$ in equation \ref{eq:pre_post_shock} and cancelling out the $v_i'$ term we arrive at our full expression for shock (equation \ref{eq:shock_expanded}):

\begin{equation}
    \mathcal{E}=|\frac{v_0-v_s}{v_0}|^2 = \left(1-\sqrt{\left(\frac{1}{2}\right)^2+\left(e_n\frac{\sqrt{3}}{2}\right)^2}\right)^2
    \label{eq:shock_expanded}
\end{equation}

In our simple Quartier model $e_n$ must be equal to zero otherwise we break the assumption of constant contact, and $\mathcal{E}$ reduces to $\frac{1}{4}$.  Thus according to equation \ref{eq:steady_states}, the critical angle which bifurcates the system into perpetual motion, or eventually stopped is $\theta^c=10.21^\circ$. 

\begin{figure}[h]
    \includegraphics[width=.45\textwidth]{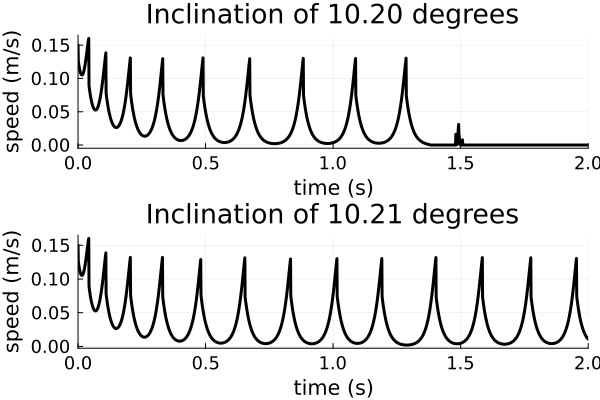}
    \caption{Velocity time series for a frictionless grain sliding over a rough inclined substrate as produced from a MFiX simulation of our frictionless Quartier system.}
    \label{fig:mfix_verify}
\end{figure}

It is worth noting that, as our system is frictionless, these governing equations (and the derivation method) are distinct from Quartier's original approach and results (Ref.~[\onlinecite{quartier_dynamics_2017}]). As a validation of our analysis, we construct a discrete element model simulation of our frictionless Quartier system.. To do this, we implemented the U.S. Department of Energy's MFiX-DEM approach for frictionless mono-disperse grains (Ref.~[\onlinecite{garg_documentation_nodate}]) .  A two-dimensional rectangular domain is created with cyclic boundary conditions for the vertical edges of the domain. One layer of grains is fixed to the base by resetting the acceleration to zero at the end of each computation cycle. A single grain is placed in the well between the first and second particle and is set to have an initial velocity sufficient to escape the gravity well. The material parameters can be set such that the friction coefficient is zero and the normal restitution coefficient is nearly zero ($10^{-20}$). Finally, the vector for gravity is inclined from vertical to reproduce the angle of inclination.  The minimum angle at which the grain can maintain steady state matches the critical angle calculated from finding the roots to equation \ref{eq:steady_states} to one-hundredth of a degree (Figure \ref{fig:mfix_verify}).

While many elements of this model are idealized, particularly the demand that $e_n=0$, we see that even in this simple form this model can explain a potential root cause for hysteresis. Namely, that the angle to maintain steady motion ($\theta=10.21$) is less than the angle needed to start moving from rest $\theta >30$. This rest angle stems from the fact that at an inclination of thirty degrees, the mobile grain is balanced right at the top of particle B (Figure \ref{fig:force_diagrams}), and at any perturbation would begin to slide on its own.

While this model is insightful for showing how simple consequences of system geometry can give rise to more complex granular behavior, it fails to capture anything resembling the thin layer stability we see manifest in real granular systems. How could it when the governing dynamics are only dependent on one lengthscale (particle size)? In order to capture this dependence of critical angle and non-dimensionalized system height, the dynamics of our system must depend on more than one lengthscale, which can be naturally accomplished through the inclusion of deformation as described in the introduction.

\section{\label{methods}Methods}
\subsection{\label{modified_quartier}The Modified Quartier Model}
Our modified Quartier model makes two changes to the traditional setup:(1) the particles are allowed to deform  and (2) we introduce a new independent variable, the number of layers of stacked particles that remain in contact above the substrate, $n$. 

\begin{figure}[h]
    \includegraphics[width=0.45\textwidth]{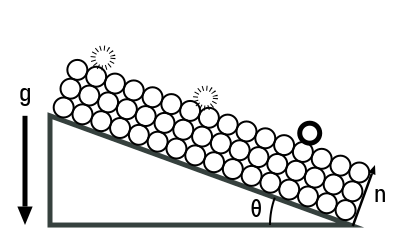}
    \caption{The modified Quartier model allows for the substrate to be a variable height of $n$ grains.}
    \label{fig:modified_quartier}
\end{figure}

By allowing for deformation, the dynamics of the system are no longer instantaneously decided in the shock, because the contact force ($F_c$) can be larger than the normal component of gravity as the grain crests the next particle, and the difference in these forces will be accommodated by deformation. Furthermore, this contact force is the only means by which our mobile grain can oppose the driving force of gravity,  and by allowing for deformation it becomes dependent on a series of interactions between compressed particles as the shock from entering the gravity well propagates down the substrate.  The length scale imposed by the compressible bed then dictates the geometric friction component of the mobile grain.  That is, the dynamics which unfold as the grain leaves the gravity well and either crests the next particle or remains trapped, are now determined by compressive chains which form between the grain and the substrate. Before we can incorporate these effects into the equations from the previous section, we must first discuss the now relevant physical parameter of stiffness, and also ensure that the time scale over which these compressive dynamics unfold are well situated within the timescale of a cresting particle. 

In many discrete element models, the normal contact force felt between two particles, our $F_c$ in figure \ref{fig:force_diagrams}, is determined by the material property of the ``normal stiffness coefficient " given by $k_n$, where the subscript is necessary to demarcate it from the stiffness of the material when sheared.  This sheared stiffness is not relevant to our frictionless model, so we drop the subscript and use ``k" and stiffness to refer to this normal stiffness.  A non-dimensional stiffness gives the rigidity ($\kappa$) and this is usually denoted by the stiffness normalized by the confining pressure. This is shown in equation \ref{eq:rigidity}

\begin{equation}
    \kappa = \frac{k}{P}
    \label{eq:rigidity}
\end{equation}

da Cruz (2005) notes that for $\kappa \ge 10,000$, there is little variation in coordination number with further increases to rigidity and the length scales for particle deformation become negligible to gaps between neighboring grains.  Providing an example for physical intuition, a bead of glass at the bottom of a 50 cm deep pile, will have a rigidity of 40,000 (Ref.~[\onlinecite{DaCruz2005}]). Our analysis will include rigidity values ranging three order of magnitude [10,000, 1,000,000], because our confining pressure for the mobile grain is determined by the gravitational body force ($9.81  \,m/s^2)$ this maps to stiffnesses of  100,000 $N/m$ and 10,000,000 $N/m$, respectively.  The lower bound of stiffness is used for the timescale calculations  as this will give us an upper bound for the timescales under consideration.

The timescale, which will dominate the Quartier model, is the time it takes for a grain to rise from particle well to crest (point b to c in figure \ref{fig:quartier_model}). This is estimated by taking the minimum change in velocity needed to escape the well and dividing by the acceleration provided form the gravitational field (equation \ref{eq:time_crest}).

\begin{equation}
    t_{c} = \frac{\Delta V}{a} = \frac{\frac{1}{4}\sqrt{2g\left(2r\left(1-sin(60)\right)\right)}cos(60)}{g}\approx 0.41 s
    \label{eq:time_crest}
\end{equation}

We use the equation for rigid particle contacts to estimate the timescale for deformation (Ref.~[\onlinecite{garg_documentation_nodate}]) . This is given by equation \ref{eq:time_rigid_contact}. 
\begin{equation}
    t_{n} = \pi\left(\frac{k_n}{m_{eff}}-\left(\frac{\gamma}{m_{eff}}\right)^2\right)^{-\frac{1}{2}}\approx 0.007s
    \label{eq:time_rigid_contact}
\end{equation}

In equation \ref{eq:time_rigid_contact} $\gamma$ is the damping coefficient (here, $\gamma = 0$), and $m_{eff}$ is the effective mass, defined as $m_{eff} =\frac{m_1m_2}{m_1+m_2}$, where $m_1$ and $m_2$ are the masses of the two colliding particles. The effective mass is used to translate binary particle contacts into a center of mass reference frame. Using our lower bound on stiffness (100,000 $N/m$ ) gives us an upper bound on time,  $t_{n}= 0.007$ seconds. Thus, these interactions are well separated, and we can expect the significant  dynamics introduced through deformation to play out while the crests (moves from point B to C in figure \ref{fig:quartier_model}). 

As mentioned at the start of this section, these compressive dynamics will modify the contact force(the singular source of friction) and ultimately the impulse imparted to the mobile grain as it leaves the particle well. No longer an instantaneous shock, the ability for the substrate to impart this impulse is dependent on a complex series of deformable collisions, whose dynamics are asymptotically limited by the dimensionless height of the substrate(n). However, if we know how the impulse imparted by these deformable collisions ($J_n$) compares to the ideal impulse of the instantaneous shock ($J_0$),  we could appropriately rescale the original post shock velocity (equation\ref{eq:pre_post_shock}) as shown below(eq \ref{eq:modified_shock}). In a  slight abuse of notation, and loose use of the word ``shock" we denote the deformation modified versions of post-shock velocity ($v_i$) and shock $(\mathcal{E})$ as functions of non-dimensionalized height $n$, while the original versions are just written as scalars.

\begin{equation}
\begin{split}
\alpha(n) & = \frac{J(n)}{J_0} \\
v_i(n) & =v_i\cdot \alpha(n) \\
\mathcal{E}(n) & =\left|\frac{v_0-v_s\cdot\alpha(n)}{v_0}\right|^2 
\end{split}
\label{eq:modified_shock}
\end{equation}

With these modified equations, we will be able to extend the tractable power of the quartier model not just to systems with deformation, but to systems with significant (because it effects system dynamics) lengthscales spanning the micro and macro-scale.

In order to actually capture this percentage change in impulse as a function of compression chain length, we need to be able to model the deformable collisions of one dimensional particles.  Quantitatively capturing the effect of the length of compression chain on the contact force, requires modifying equation \ref{eq:effective_k} to be a description of massive separable springs, which, in turn, requires the modeling of 1D deformable particle collisions. 

\subsection{\label{deform_collisions}Modeling Deformable Particle Collisions}
The equation for springs in series (equation \ref{eq:effective_k}) can be seen as a simple constitutive equation for a system of springs which succinctly captures an emergent effect.  This effect is implicitly captured if each spring is analyzed individually. Similarly, the bulk effect for particles compressing in series can be captured by resolving particle-particle interactions in DEM's such as LAMMPS and MFiX. However, what is missing is a description of the emergent effect, a granular equivalent to equation \ref{eq:effective_k}; such a description is necessary to complete equation \ref{eq:modified_shock} and close our analysis of the modified quartier model.  This is necessary to be able to say anything prescriptive about phenomenon which emerge from these compressions in series, however, rather than using existing DEM approaches to capture this effect, we choose to model deformable collisions directly.

Existing DEMs resolve the contact force between two particles from a particle-overlap (Ref.~[\onlinecite{garg_documentation_nodate}]), and while this scheme will capture the reduction in work done between two particles as work is passed further down the chain of contacts, it does not capture the minuscule gaps in contact and translations that would occur in real deformation. Because it is the lengthscale of contact which sets the reduction in effective stiffness, it's evolution in time is important.  To fully resolve the length of contacts which form in simple one-dimensional collisions, the dynamics must be modeled from deformation.  The methodology for achieving this is most clearly illustrated in the two particle example, where one deformable particle slides into another deformable particle initially at rest. The supplemental material includes videos of these simulations for two and three particle systems, which succinctly illustrates the results laid out over the next several paragraphs and provides intuition for the complexity in the structures of contacts introduced just by going from two to three particles. 

Deformation is linked to inter-particular forces through the linear dash-pot model (equation \ref{eq:linear_dashpot}), where $k$ represents the material property of stiffness and $\Delta x$ represents the lengthscale of deformation, which is assumed to be axisymmetric.

\begin{equation}
    F = k \cdot \Delta x \rightarrow a = \frac{k}{m}\Delta x
    \label{eq:linear_dashpot}
\end{equation}

If $r_0$ represents the relaxed radius of our particle, then the distance between the two deformed particles can be represented as the sum of the deformed radii. By newtons third law and equation \ref{eq:linear_dashpot} these deformations must be equivalent.
\begin{equation*}
    x_2-x_1 = \underbrace{\frac{(2r_0-\Delta x_1)}{2}}_{\text{particle 1}}+\underbrace{\frac{(2r_0-\Delta x_2)}{2}}_{\text{particle 2}}=2r_0-\Delta x
    \label{eq:pos2deform}
\end{equation*}

This allows us to write the equations of motion as a function of particle position:
\begin{equation}
    a_1 = \frac{k}{m}\left(2r_0-\left(x_2-x_1\right)\right) \,\,\,\,\,\,\,\,\,\, a_2 = -\frac{k}{m}\left(2r_0-\left(x_2-x_1\right)\right)
    \label{eq:pos2accel}
\end{equation}

\begin{figure*}
    \includegraphics[width=.80\textwidth]{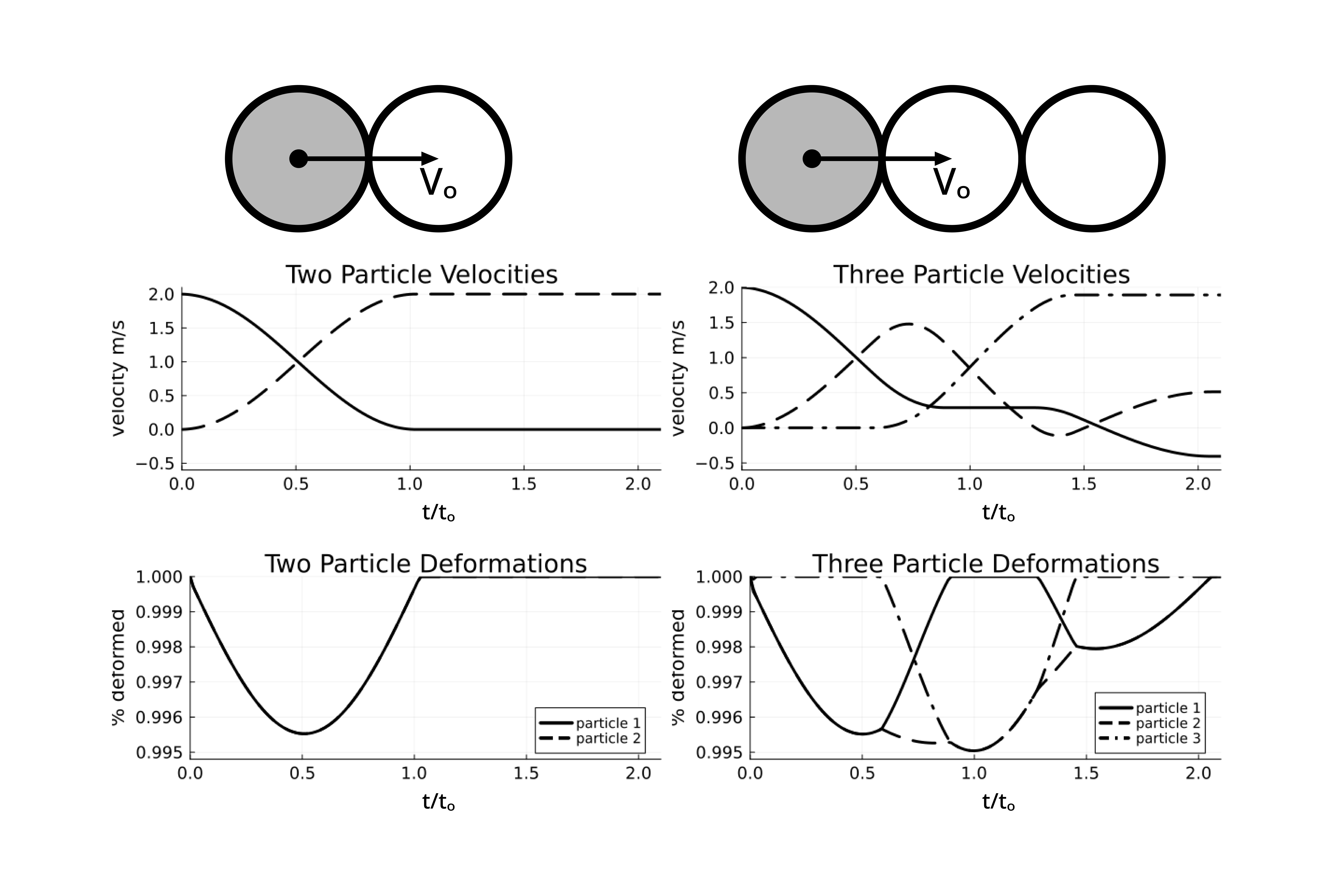}
    \caption{Velocity and deformation outputs from the deformable collision model. The left and right side depict two and three particle results respectively. }
    \label{fig:2p_3p_RK4}
\end{figure*}

Even for only a two particle system, these equations have no analytical solution. However, the link between acceleration and particle position makes it easy to implement a numerical solution,using the classic 4th order Runge-Kutta method (RK4). The results for this model run for particles of rigidity $\kappa=100,000$ (our median rigidity) are summarized in the left hand side of figure \ref{fig:2p_3p_RK4}, where we recover the expected total exchange of momentum (mobile grain ends with zero velocity).  

\begin{figure}

    \includegraphics[width=.45\textwidth]{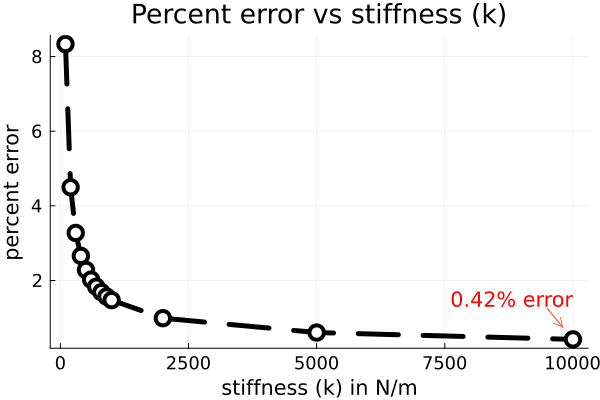}
    \caption{The percent error in deformations for a two particle system as a function of stiffness (k). Percent error is calculated between the maximum  deformation predicted from conservation equations and the maximum deformation observed in our model. The minimum stiffness used in our simulations is 100,000 N/m.}
    \label{fig:k_error}
\end{figure}

With two particles, conservation of momentum and energy allow us to solve for the expected deformation and velocities,  and we can calculate an error of our numerical approximations which is shown in figure \ref{fig:k_error} .  The error appears to significantly decrease with particle stiffness, becoming less than half-a-percent by $k=10,000 N/m$, which is  an order of magnitude smaller than the lowest stiffnesses in our numerical experiments. 

As we increase system size, we increase the chain of particles initially at rest and touching. Thus a three particle system has one active particle colliding into a chain of two at-rest particles aligned through their center of mass.  In systems larger than two particles, there are no longer unique solutions to momentum and energy conservation. Furthermore, the deformation of the particles is no longer symmetric. This  is highlighted in the right-hand-side of figure \ref{fig:2p_3p_RK4}, and can be seen at the moment that the first two particles collide with the third, before this occurs particles one and two have equal deformation, as soon as they form a three particle chain, the second particle will be additionally deformed by its contact with the third particle, which will in turn be the least deformed particle in the chain. However, the constraint of equal deformations can be replaced with the constraint that partial deformations must be equivalent. In our three particle example this states, that the component of deformation for particle two caused by particle three must be equal to the deformation of particle three. Thus, modeling n-particle deformable collisions becomes an exercise in recovering unique boundary deformations from center of masses.  A full description of this procedure is covered in appendix \ref{sec:recover_boundaries}, but we highlight the key points below.

Given the center of masses for an n particle system, we can guarantee that the two closest centers must be in contact.  This provides sufficient information to identify which particles form a compression chain. Once we know which particles  participate in a compression chain, the constraint of equal partial compressions can be used to recover the boundaries. The equation for recovering the boundaries of n particles in a compression chain is given by equation \ref{eq:boundary_solution} and equation \ref{eq:boundary_recurrence}.
\begin{equation}
    b_0= \frac{1}{n}\left(\left(-1\right)^nr_0+\Sigma_{i=1}^n\left((-1)^{i+1}\left(2(n-i)+1\right)x_i\right)\right)
    \label{eq:boundary_solution}
\end{equation}

\begin{equation}
    b_n = x_n +(x_n-b_{n-1})
    \label{eq:boundary_recurrence}
\end{equation}

Equation \ref{eq:boundary_solution} finds the leftmost boundary ($b_0$) from the center of masses ($x_i$'s) of a $n$ particle chain, and then equation \ref{eq:boundary_recurrence} leverages the assumption of axisymmetric deformation to create a recurrence relation for boundaries of particle's in a compression chain (all boundaries can be found from one, but finding the first one is not trivial). Once these boundaries are known, so too are the deformations, and we can implement our RK4 for a system of n particles. 

\textbf{In summary, there is enough information encoded in the particle accelerations, positions, and physical constraints to uniquely reconstruct particle contacts and deformations.}

Once the deformation and contacts of each particle are recoverable from particle center of mass, an analogous Rk4 model can be used to simulate systems of arbitrary size ( we simulate up to 20 particles). The right hand side of figure \ref{fig:2p_3p_RK4} illustrates how quickly the dynamics change with the addition of just one particle.  Two characteristic changes in system behavior include (1) an incomplete transfer of momentum, shown in the negative final velocity of the first particle, and (2) an increase in total deformation, and thus energy stored, for the three particle system. From the active particle's point of view both of these changes are analogous to colliding with one particle of reduced stiffness, manifesting the exact weakening effect we were expecting. Furthermore, while the change in total deformation is small, the stiffness dictates this is a significant difference in energy storage. 

The final two additions we must make to our deformable particle model before we can use it to calculate impulses and ultimately $\alpha(n)$: (1) we fix the $n^{\text{th}}$particle in space (it represents the particle at the bottom of our substrate), and (2) we add an optional gravitational field in the direction of motion.  Thus, our model can capture horizontal frictionless deformable collision (air hockey table) and vertical (dropping a grain onto a tower of stationary particles). With this model developed, we next examine implications for the contact force of the bed-modified Quartier model.

\section{\label{results}Results}
Our results section begins with an analysis to extract the impulses felt by the mobile grain in a one dimensional deformable collision model, and ends with scaling results for thin layer stability observed in real granular systems. We begin this section with some scaffolding to help the reader remember how one result builds into the other. 
\begin{itemize}
    \item \textbf{Result 1:} By modeling the dynamics of deformable grains, we can capture a modified version of the effective spring constant for massive separable springs, which is necessary for capturing the effect of compressed particles in series. 
    \item \textbf{Result 2:} From these models, we calculate the impulses felt by a mobile grain colliding into a chain of $n$ particles.
    \item \textbf{Result 3:}  These impulses can be used to describe the contact forces of a mobile grain in our modified Quartier model, culminating in our final result, the critical angle of inclination as a function of system thickness. 
\end{itemize}

\begin{figure}[h]
    \includegraphics[width=.45\textwidth]{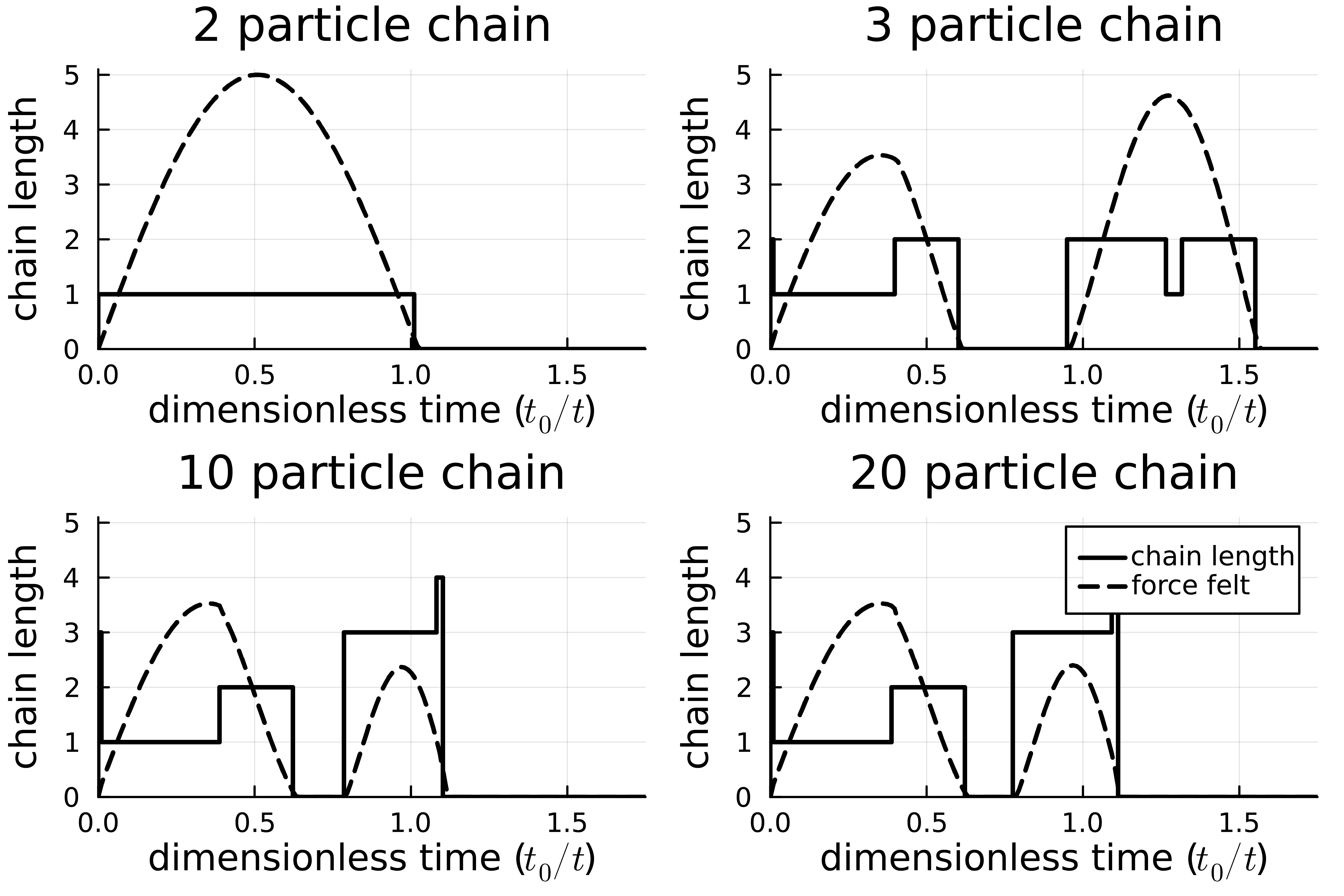}
    \caption{The dashed lines trace out the magnitude of forces felt by a mobile grain, under the influence of gravity, as it collides into a vertical chain of particles of various length. Forces are normalized such that proportional comparisons between various chain lengths is accurate. The solid line indicates the number of additional particles currently deforming in series with the mobile grain (chain length). Time is non-dimensionalised such that the two particle interaction is one unit of dimensionless time. }
    \label{fig:results_g}
\end{figure}

Our first set of results elucidates the evolution of contact structures throughout the collision of one particle into a static chain of length $n$, and how these structures, in turn, affect the forces felt by the original particle. To emphasize the link to the Quartier model, the particle originally imbued with a velocity will be referred to as the grain, and the rest of the static chain will simply be particles. The direction of motion is aligned with the gravitational field, and the last particle is fixed in space, so these simulations can be thought of as dropping a grain onto a tower of $n$ grains, perfectly aligned along center of mass. Simulations are run for $n$ ranging from one to nineteen (two to twenty particles total).  We highlight results for the two, three, ten, and twenty particle cases  in figure \ref{fig:results_g}. We selected two and three particle scenarios to emphasize the immediate and significant dynamic changes, and the ten and twenty particle scenarios demonstrate the eventual convergence of behavior. 

\begin{figure}[h]
    \includegraphics[width=.45\textwidth]{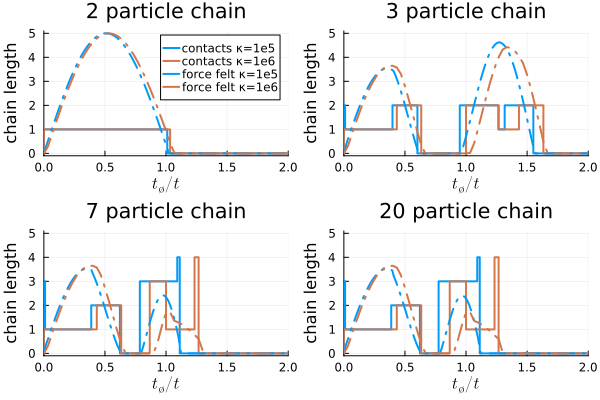}
    \caption{An illustration of how our non-dimensionalization collapses results from simulations run with rigidities seperated by an order of magnitude.}
    \label{fig:contacts_k_compare}
\end{figure}

These plots trace the absolute value of a normalized force felt by the mobile grain and the length of its resulting compression chain through dimensionless time. The force is normalized to the largest force felt across all 19 simulations, and then scaled up by the contact structure number to plot nicely on the structure plot. This preserves the value of relative comparison between plots (if the peak of one simulation looks twice as large as another simulations peak - it is). Additionally the accelerations due to gravity are subtracted out, because we just want to visualize the forces caused by particle contacts.  The time for each simulation is normalized for the duration of contact time for the two particle simulation ($t_o$). Facilitating comparisons between simulations run at different rigidites. Figure \ref{fig:contacts_k_compare} shows that this non-dimensionalization nicely collapses data produced from simulations run with stiffnesses varying by an order of magnitude. We walk through the three particle simulation of figure \ref{fig:results_g} as an example for interpreting these plots. 

At the first time step, the grain is in contact with the second particle, which is in contact with the third particle. Only the first particle has a velocity. At the second time step, the first and second particles deform, breaking contact with the third particle. These two particles continue to deform and move until a non-dimensional time of $\sim$0.5. At this point, the deformed particles collide with the third particle, forming a three-particle chain. This is shown by the solid line stepping from one (the mobile grain's compression chain includes one other particle) to two (the chain includes two other particles, making three compressed particles in total). As the mobile grain rebounds and the second and third particles continue to deform, there is a period when the mobile grain loses contact with the rest of the chain. This is indicated by both the contact line and forces-felt dashed-line dropping to zero. At one unit of dimensionless time, the decompressing second and third particles re-initiate contact with the mobile grain. We see this in the first particle registering forces again (dashed-line rises from zero), and in the contact line stepping up to two.  These interactions continue until about one-and-a-half units of dimensionless time, when the mobile grain is ejected back up away from the pile.

Essential in these dynamics are the alignment between characteristic changes in these force curves with change in contact structure, which ties back in with the effective force having a series like dependence on number of particles in the chain \ref{eq:effective_k}. When we calculate our second result, we will effectively be calculating the impulses for systems of massive separable springs in series, as mentioned in the introduction.

\begin{figure}[h]
     \includegraphics[width=0.45\textwidth]{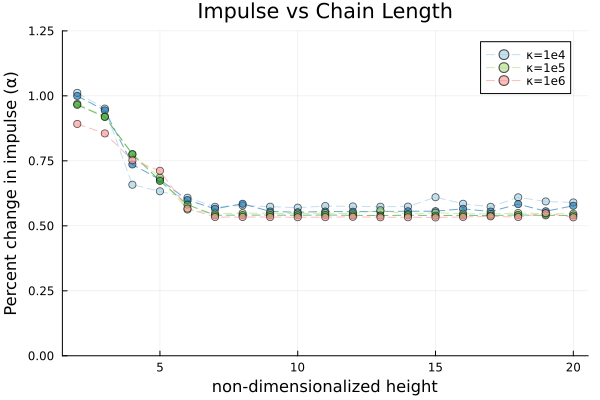}
     \caption{Blue, green, and red curves show impulse calculations for simulations run at rigidities of $10^4$, $10^5$, and $10^6$ respectively. Lighter and darker shades of each color represent initial velocities of 2 m/s and 4 m/s respectively. }
     \label{fig:impulses}
 \end{figure}

Our second result is the impulse, which is calculated from each of these simulations. Once again, these results expressed in a dimensionless quantity of percent impulse change, or $\alpha(n)$ (eq \ref{eq:modified_shock}). This variable is written as a function of chain length $n$ to emphasize the dependence. Similar to our dimensionless time, this dimensionless impulse is normalized to the expected result in the two particle scenarios. In both of these cases, the two particle simulation represents the null hypothesis that these quantities should only be determined by interactions between direct neighbors.  To show the generality of these results, we calculated $\alpha(n)$ for six simulations with rigidity spanning three orders of magnitude ($10^4$ to $10^6$) and with two different initial velocities for the mobile grain (2 and 4 m/s),(Figure \ref{fig:impulses}).

The most difficult part in extracting impulses from these plots is identifying the appropriate timescale over which to preform the integration. This is covered comprehensively in appendix \ref{sec:time_scale}, but in summary, the impulse is calculated over a timescale determined by the maximum compression of the particle chain.

As expected, the impulses asymptotically decrease with chain length. This asymptotic behavior is due to the limitations small system sizes place on the ability to form compressive chains. However, chain lengths plateau at around ten particle contacts, so increasing system size beyond that has little to no-effect. The x-axis is labeled as non-dimensional height, to emphasize the link between the impulse calculated for an $n$-chain one dimensional system, and the capacity for a modified Quartier model of height $n$ to resist the driving force of gravity. In other words, we are seeing the asymptotic weakening of a system's geometric friction with increased system size. 

\begin{figure}[h]
    \includegraphics[width=.45\textwidth]{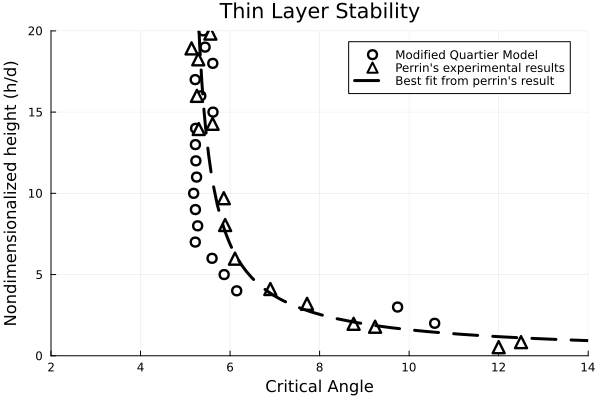}
    \caption{Circles are our analytical results for critical angle as a function of non-dimensionalized height (eq \ref{eq:steady_states}). Triangles and dashed line are Perrin's experimental results and line of best fit}
    \label{fig:our_TLS}
\end{figure}

This is the general behavior for a mechanism which is capable of generating thin layer stability, and we can use our formal analysis of the modified Quartier model to check the scaling produced. To do this, we calculate our critical angle (equation \ref{eq:steady_states}) as dependent on the modified shock term (equation \ref{eq:modified_shock}). This recovers the experimentally observed behavior of thin layer stability, as shown in figure \ref{fig:our_TLS}.

These results would indicate that the physical mechanism of compression chains is both necessary and sufficient to explain the emergence of thin layer stability for frictionless, mono-disperse, inclined granular flow. Furthermore, the fundamental nature of the granular system we are analyzing would suggest that compression chains play a necessary, if not sufficient, role in setting meso-scale dynamics in any granular systems where the timescale of compression in series is less than that of particle re-arrangement.  Thin layer stability provided an excellent framework within which we could build an exacting  and testable hypothesis. Now that it's existence has been thoroughly demonstrated, it is worth using a less exacting gaze and considering some of the more general implications of softness in rigid grains, extending these insights into less simplified granular systems. 

\section{\label{sec:discussion}Discussion}

Our findings reveal the significant role of softness in granular systems, particularly through the mechanism of compression chains. By allowing for particle deformation, we have demonstrated that the resulting compression in series influences local coefficients of effective friction, directly affecting the stability of granular flows. However, in most granular systems, compression in series will not manifest in such ideal configurations.

\begin{figure}
    \centering
    \includegraphics[width=0.5\linewidth]{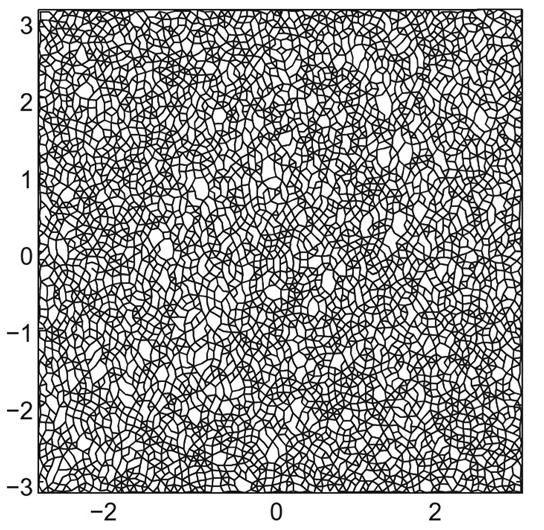}
    \caption{Visualization of contact network formed from two-dimensional polydisperse disks as shown in Popadopoulous 2018[\onlinecite{Papadopoulos_2018}]}
    \label{fig:2d_contacts}
\end{figure}

Figure \ref{fig:2d_contacts} illustrates the contact network for a two-dimensional packing of polydisperse disks as shown in Papadopoulos (2018) (Ref.~[\onlinecite{Papadopoulos_2018}]). As with any complex contact structure, they are no longer clean, straight lines,  but these structures still facilitate compression in series.  In lieu of performing a full decomposition of contact forces based off of interior angles formed from the contact network, as would be required by our original approach,  we take a simpler first order approach towards describing how softness effects the energy storage capacity of these contact structures:a behavior which provides a natural bridge for the gap between microscopic particle interactions and macroscopic flow behaviors.

 Under the linear-dashpot model, a particle with rigidity, $k$ subject to force $F$, will be compressed by some amount  $c_1=\Delta x$ .  Two of these  particles compressed in series have an effective rigidity $k'=k/2$(per linear-dashpot assumption), and if subject to the same force will  experience twice the compression, $c_2 = 2c_1$, as shown below:

\begin{equation}
F = k'\cdot c_2 = \frac{k}{2}\cdot 2c_1=k\cdot c_1 \nonumber
\end{equation}

This also means the two particle system will store twice the energy, because the potential energy only depends linearly on $k$ yet quadratically on $\Delta x$ . This is shown below where the subscript indicates the systems particle number, and each system is subject to the same force. Thus,  $U_2$ represents the potential energy of a two particle system. 

\begin{equation}
    U_1=\frac{1}{2}k c_1^2\nonumber
\end{equation}
\begin{equation}
    U_2=\frac{1}{2}k'c_2^2=\frac{1}{2}(\frac{1}{2}k)(2c_1)^2=\frac{1}{4}\cdot4c_1^2=2U_1\nonumber
\end{equation}

In general we have  $U_n=nU_1$. Not surprisingly, the more particles which are able to compress in series, the more efficient that contact structure will be at storing potential energies. This is verified in our numerical models and can be seen by comparing the bottom plots in figure \ref{fig:2p_3p_RK4}. The three particle system achieves a larger total deformation.  This increased capacity to store potential energy can also be viewed as an increased effectiveness at damping velocity fluctuations,  an effect which grows like the square-root of the compression chain lengthscale. This is seen in the kinetic energy which can be accommodated by a spring system of length $n$ (equation \ref{eq:energy_scaling})

\begin{equation}
    KE = \frac{1}{2}mv^2 = U_{n}=nU_1=n\left(\frac{1}{2}k\cdot c_n\right)\rightarrow v \propto \sqrt{n}
    \label{eq:energy_scaling}
\end{equation}

\begin{figure}[h]
    \includegraphics[width=.45\textwidth]{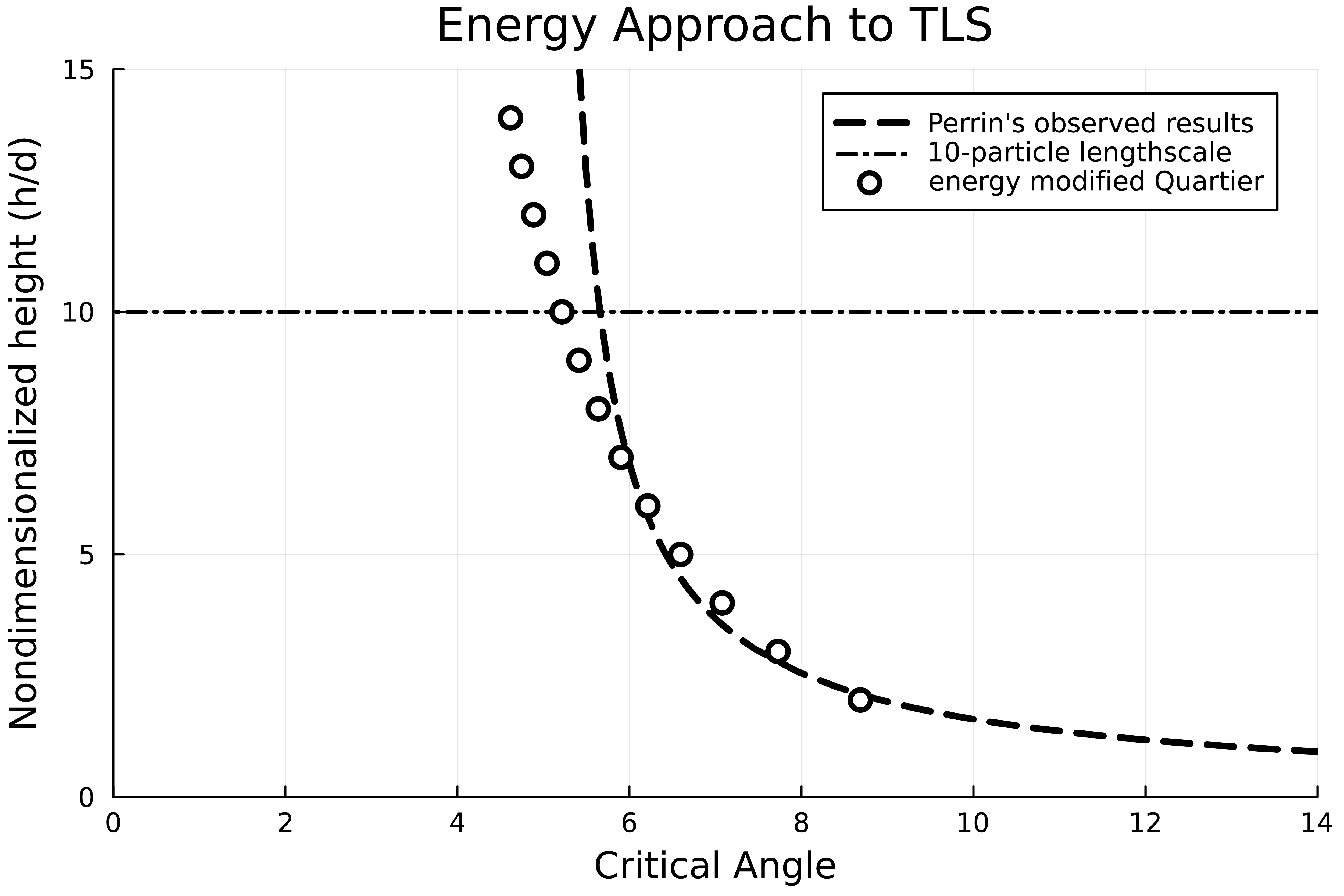}
    \caption{Scaling of critical angle produced by modifying the effective shock (eq\ref{eq:modified_shock}) according to the results of our energy budget analysis. The horizontal dashed line represents the 10-particle lengthscale cut off (a common threshold for mesoscale). }
    \label{fig:geom_scaling}
\end{figure}

Revisiting figure \ref{fig:2d_contacts}, the variations in contact geometry and structure can now be linked to a rich field of information describing local spatial variations in ability to damp velocity fluctuations.  Put another way, compression through these contact structures introduce a non-local lengthscale (length of compression) which is determining the diffusion of granular temperature (a description of particle velocity fluctuations).  Within the context of our original thin layer stability argument,  the legnthscales in the contact network are controlled by the height of the fixed bed. As the lengthscale of the fixed bed increases so too does damping efficiency due to deformability (equation \ref{eq:energy_scaling}), which leads to a strong reduction in the contact structures ability to impose geometric friction.  As a quick check we can see how this general energy approach could similarly (albeit less explicitly) capture the emergent behavior of thin layer stability. By modifying the velocity loss between energy traps to have the same proportionality as exhibited in equation \ref{eq:energy_scaling}, we can produce figure \ref{fig:geom_scaling}, which recovers thin layer stability up to a lengthscale of approximately ten particles.  While the asymptotic behavior is not recovered,  our results do not show any particle contacts over ten-particles forming, even for the twenty particle chain. Furthermore this ten-particle lengthscale shows up frequently in the literature as a way to delimit the mesoscale(Ref.~[\onlinecite{PhysRevE.89.012203}]).  In this manner, the asymptotic behavior would be set by the fabrics inability to form many strong contacts longer than 10 particles.

These results underscore the necessity of incorporating deformability into models of granular rheology. By doing so, we can begin to move beyond phenomenological descriptions and towards a more precise understanding of how microscale behavior constructs mesoscale and non-local effects.

\section{\label{conclusion}Conclusion}
By removing the constraint of infinite rigidity, we create space for a mechanism to manifest which uses the compounding effect of compression through chains of particles to modify the effective rigidity in the neighborhoods of its terminal points. While forces ranging an order of magnitude produce deformational differences less than 0.000001\%  in particles with rigidity of $10^6$, allowing for deformation and the formation of compression chains introduces a dynamic which can nearly halve the impulse generated from contact forces (Figure \ref{fig:impulses}). These contact forces, as emphasized in the Quartier model, can play an important role in determining the geometric friction of the system. Accounting for this phenomenon in the Quartier model allows for the construction of a prescriptive description of the critical stresses in inclined, planar, and frictionless granular flow which recovers the hereto unexplained scale-dependent and non-local dependence on system height (thin layer stability). More generally, the number of simultaneously compressed particles introduces an important new lengthscale (attached to an understood kinematic effect), which links a mesoscale structure to local rheologic ability to dampen velocity fluctuations. This elucidates the natural role of deformability as a bridge between microscopic particle interactions and macroscopic flow behaviors. The propensity for a system to form compression chains and their respective efficacy will be set by properties of the contact network, opening up many fascinating paths of inquiry into the geometry of structures such as figure \ref{fig:2d_contacts} . Ultimately, we wish to expose the rich ecosystem of new interactions which can only arise from relaxing the assumptions of infinite rigidity and to motivate the necessity of grappling with deformation  directly  if we want to move beyond phenomenological descriptions for mesoscale and non-local effects. 

\section{\label{ack}Acknowledgments}
Financial support to C.H and J.D. provided by NSF EAR grants 1926025 and 1949219. E.C.P.B. was supported by UKRI with the NERC-IRF (NE/V014242/1).  

\appendix

\section*{Appendix}
\appendix
\section{\label{sec:recover_boundaries}Resolving Boundaries from Center of Mass}

Recovering the boundaries for a one-dimensional system of deformable particles based only on center of masses can be framed by two questions; Can we identify which particles are compressed in chains (i.e. where are the gaps), and then, within each chain of compressed particles, can we recover deformed boundaries? We start with the second question. Throughout this section assume that numerical and alphabetical indexing order a particles position from left to right. Thus particle $i$ or $1$ is left-adjacent to particle $j$ or $2$.

\subsection{Recovering boundaries within a compressed chain} 
If we have a compressed chain of $n$ particles and the positions of their center of masses $\{x_1,x_2,...,x_n\}$, then there exist $n+1$ boundaries $\{b_0,b_1,...,b_n\}$,  and without loss of generality we make $b_0$ the left most boundary in the chain. These boundaries are recoverable with two constraints: (1) axisymmetry of deformation and (2) Newton's third law.  By axisymmetric deformation we can write a recurrent relationship for the particle boundaries.

\begin{equation}
\begin{gathered}
    b_0 \\
    b_1 = x_1 + (x_1 -b_0)\\
    b_2 = x_2 + (x_2-b_1) \\
    b_3 = x_3 + (x_3-b_2) \\
    \vdots\\
    b_n = x_n +(x_n-b_{n-1})
    \label{eq:boundary_recur}
\end{gathered}
\end{equation}

Simply put, the above relation states that the right boundary of the first particle must be the same distance from the center of mass as the left boundary, and that this right boundary is the left boundary of the next particle, and so forth...(remember we start by assuming these particles are in contact and part of a compressive chain). This relationship allows for every boundary to be written as the particle's center of masses and the left most boundary.
\begin{equation}
\begin{gathered}
    b_0 \\
    b_1 = 2x_1 - b_0 \\
    b_2 = 2x_2-2x_1+b_0 \\
    b_3 = 2x_3 -2x_2 +2x_1 -b_0 \\
    \vdots \\
    b_n = \Sigma_{i=1}^n \left(2x_i\left({-1}\right)^{n+1} +b_0\left(-1\right)^{n}\right) 
    \label{boundary_recur_simp}
\end{gathered}
\end{equation}

Since the total compression of the chain is known from information about the center of mass positions ($x_n-x_1$) so too is the length of the chain, which provides us enough information to find all the boundaries for odd chains of compression. We provide an example with a three particle chain, by subtracting $b_0$ from both sides of the equation above:
\begin{equation*}
    \underbrace{b_3-b_0}_{\text{length}} = 2x_3-2x_2+2x_1-2b_0
\end{equation*}
All of the positions are known, therefore we can solve for $b_0$ and use the recurrence relationship to find all particle boundaries. Interestingly, this is not the case for even chains of compression because the $b_0$ term on the right hand side cancels out.  This surprising result makes more sense when considering a chain of two particles, which can have identical centers of mass positions and chain length but different boundaries by swapping deformation of each particle; odd particles break this symmetry. However, Newton's third  law removes this degree of freedom (in a two particle chain the deformation of the two particles must be identical).

We incorporate Newton's third law by linking the distance between the the position and boundaries to the compression in our linear dashpot model. The total compression of partilce $i$ is given by the difference between its relaxed length ($2r_0$) and its deformed length ($(b_{i}-x_i)+(x_i-b_{i-1})$). Once again using the assumption of axisymetric deformation this can be simplified into equation \ref{eq:boundary_2_deform}
\begin{equation}
    \Delta x_i = 2\cdot(r_0-(x_i-b_{i-1}))
    \label{eq:boundary_2_deform}
\end{equation}
The total compression of some particle $i$ will have two contributing factors, a  partial compression from its left  and right neighbors, represented as $c_{(i-1,i)}$ and $c_{(i,i+1)}$ respectively (equation \ref{eq:partial_comp})
\begin{equation}
\Delta x_i = c_{(i-1,i)} + c_{(i,i+1)}
    \label{eq:partial_comp}
\end{equation}
The first and last particle of the chain have the property that their total compression is equivalent to their partial compression:
\begin{equation}
    \begin{gathered}
        \Delta x_1 = c_{(1,2)}\\
        \Delta x_n = c_{(n-1,n)}
        \label{eq:endmembers}
    \end{gathered}
\end{equation}

It is worth noting that the axisymmetric assumption for compression guarantees that $b_{i}-x_i=x_i-b_{i-1}$ not that $c_{(i-1,i)}=c_{(i,i+1)}$. In other words, while the total deformation is  symmetric the contributions from the left and right neighboring particles to this total deformation do not have to be equal. We can now construct a similar recurrence relationship to equation \ref{eq:boundary_recur}, but using compression instead. By Newton's third law, the deformation of the first particle must be equal to the deformation of the second particle minus the partial compress caused by the third particle. However the partial compression of the second and third particle must be equal to the total compression of the third particle minus its partial compression from the fourth, and so on...

\begin{equation}
    \begin{gathered}
        \Delta x_1 = \Delta x_2 - c_{(2,3)} \\
        \Delta x_1 = \Delta x_2 - (\Delta x_3 - c_{(3,4)}) \\
        \Delta x_1 = \Delta x_2 - \left(\Delta x_3 - \left(\Delta x_4 - c_{(4,5)}\right)\right)\\
        \vdots \\
        \Delta x_1 = \Delta x_2 -\left(\Delta x_3 - \left(\Delta x_4 - ...\left(\Delta x_{n-1} -\Delta x_n\right)\right)...\right)
    \end{gathered}
    \label{eq:deform_recur}
\end{equation}
The last partial compressive term can just be represented as a total compression by equation \ref{eq:endmembers} .  Using equation \ref{eq:boundary_2_deform} we can rewrite all of the $\Delta x_i$'s  in terms of boundaries, and then using equation \ref{eq:boundary_recur}, we can rewrite all of these boundaries in terms of $b_0$ and solve!  An example is given below for a chain of four particles:
\begin{equation*}
    \begin{gathered}
        \Delta x_1 = \Delta x_2 -\Delta x_3+\Delta x_4 \\
        2(r_0-(x_1-b_0))=2((r_0-(x_2-b_1))-...\\
        (r_0-(x_3-b_2))+(r_0-(x_4-b_3)) \\
        r_0-x_1+b_0=r_0-x_2+2x_1-2b_0-r_0+x_3-...\\
        (2x_2-2x_1+b_0)+r_0-x_4+2x_3-2x_2+2x_1-b_0 \\
        b_0 =\frac{1}{4}(-x_4+3x_3-5x_2+7x_1)
    \end{gathered}
\end{equation*}

From this boundary we can find all of the others and we have achieved the desired result of recovering all deformations from just center of mass.  The pattern for the general solution is already clear in the solution for just 4 particles (alternating sum of odd coefficients). In general the solution for $b_0$ is given by equation \ref{eq:boundary_solution_A}

\begin{equation}
    b_0= \frac{1}{n}\left(\left(-1\right)^nr_0+\Sigma_{i=1}^n\left((-1)^{i+1}\left(2(n-i)+1\right)x_i\right)\right)
    \label{eq:boundary_solution_A}
\end{equation}

We can now recover the boundaries from chains of compressed particles, what remains is to identify the chains. 

\subsection{Identifying chains of compressed particles from center of mass positions}
As mentioned above, once we know which sets of particles are in contact, we can uniquely identify particle boundaries, however identifying which sets of particles form a compressive chain from only center of mass data is quite challenging.  There are many different ways a system of particles could be grouped into sets of compressed chains which satisfy both the assumptions of axisymmetry and Newton's third law.  In order to uniquely identify the chains of compressed particles from center of mass position information alone, we add an additional constraint shown in equation \ref{eq:max_compression} .  

\begin{equation}
    c_{(i,j)} \le 2r_0-\Delta x_i
    \label{eq:max_compression}
\end{equation}

One way to interpret this inequality is that given the  positions of particle $i$ and $j$, the boundary of particle $j$ will be as compressed as possible if particle $i$ has no further particles to its left. Thus, it is not necessarily a new constraint so much as a useful inequality which stems directly from newtons third law. It is useful because it allows us to find a threshold for ``closeness" beyond which we can guarantee particle contact.  If we know particle $k$ is in contact with some particle $l$, we can make the same assumption about particle $k$'s partial compression and if the distance between particle $j$ and $k$'s center of mass is smaller than the some of their maximal compressive distance, particle $j$ can be guaranteed to be a part of particle $k$'s compressive chain.  In this way the chains are built in an order of ``these particles at least must be in contact". Particles that are part of a chain might be missed in a first sweep but they will be picked up in a later sweep because when encountered again, more information is known about the contacts of the particles on the other side (these would have been established in a previous sweep).  All that remains is to find a starting point. Given a set of particle positions and our two assumptions which two particles , if any, are in contact.

Not surprisingly the particles whose center of masses are closest are guaranteed to be in contact. The conceptual argument is fairly simple; in order for two particles to be sufficiently deformed that they are no longer in contact, their positions to their neighboring particles must be even closer. However for the sake of robustness we present a proof by contradiction using our favorite inequality (equation \ref{eq:max_compression}). 

Assume to the contrary that particles $x_i$ and $x_j$ have the minimum distance between their centers, yet are not in contact. Thus, the distance between particles $x_i$ and $x_j$ must be greater than the sum of their radii:

\begin{equation}
    x_j-x_i > \frac{2r_0-\Delta x_i}{2}+\frac{2r_0-\Delta x_j}{2}
    \label{eq:cont_cond}
\end{equation}

The deformations of particle $x_i$ and $x_j$ do not have to be equal, however we can add the minimum of the two twice, and maintain the inequality. Without loss of generality, assume that $\Delta x_i = min\{\Delta x_i, \Delta x_j\}$. Thus equation \ref{eq:cont_cond} simplifies to:

\begin{equation}
    x_j-x_i > 2r_0-\Delta x_i
    \label{eq:cont_cond_simp_1}
\end{equation}
Now we can use our important inequality. The deformation of particle $x_i$ is maximized (boundary is minimized) when $x_h$ is not compressed by any particle to it's left. By assumption, particles $x_i$ and $x_j$ are not in contact and therefore inequality can be preserved and the deformation can be written in the simple two particle case, that is we can write $\Delta x_i$ as $\left(2r_0-\left(x_i-x_h\right)\right)$:

\begin{equation}
    x_j-x_i > 2r_0-\left(2r_0-\left(x_i-x_h\right)\right) = x_i-x_h
\end{equation}
However this means that the distance between particles $x_i$ and $x_j$ are not the minimum and we have our contradiction.

\section{\label{sec:time_scale}Picking a time Scale for Impulse Calculations}

Implicit in our results is the definition of a timescale over which we calculate our impulses for each simulation.  This question of choosing a timescale is important for describing the different phases of interactions which arise in deformable particle collisions. For our simple one dimensional collision model, we begin by creating two aspirational distinctions,  mobile  grain-neighbor interactions and mobile grain-chain interactions. Mobile grain-neighbor interactions take place at a timescale which captures the initial interactions between the mobile grain and the first particle in the chain. These dynamics will still be influenced by contacts beyond one particle lengthscale, but the focus is on this initial exchange of momentum. However, when the contacts extend beyond one particle lengthscale (when we have a series of deformed particles) some of the mobile grain's energy will be transferred down the chain. Not only does this reduce the energy in the mobile grain's immediate neighbor interaction, it also creates dynamics which will be resolved with the mobile grain at a later time. These dynamics include gravitational acceleration into gaps created by the elastic propagation of deformation, and the rebounding of the initial momentum. The dynamics which resolve on this timescale are referred to as mobile grain-chain interactions.

\begin{figure}[h]
     \includegraphics[width=0.45\textwidth]{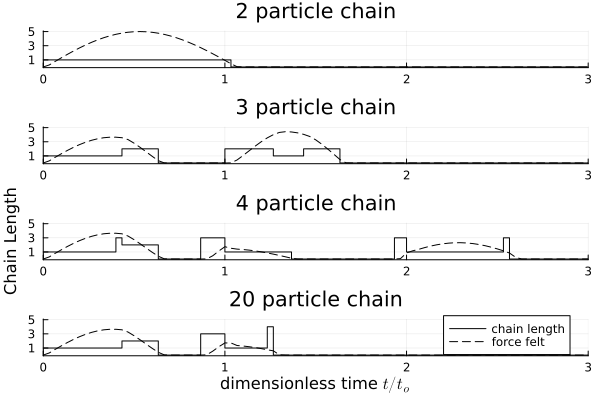}
     \caption{A clear separation of mobile grain-particle and mobile grain-chain interactions develops as chain length increases. It arises similar to a separation of constructive interference. }
     \label{fig:contact_seperation}
 \end{figure}

In the two particle simulation, mobile grain-neighbor and mobile grain-chain interactions are synonymous. In the simulations with longer chains (greater than ten particles)  these interactions are well separated. However, in the transition these interactions overlap, forming a constructive interference which makes identifying a consistent and physically intuitive timescale for the characteristic impulse difficult. We attempt to illustrate the phenomenon in figure \ref{fig:contact_seperation}. In the two particle simulation, there is no separation between the mobile grain-neighbor and chain interactions. In the twenty particle simulation, we see the initial  interactions well isolated, and the interactions which occur up to $t/t_0$=1.5 seem to be the limiting expression of the mobile grain neighbor-interactions in long compression chains. However, in the four particle simulation, we see an example of these  mobile grain-neighbor interactions (third hump) occurring on timescales which are likely still influencing the particle-neighbor interactions. The three particle case is the most extreme example, with the mobile grain-chain interaction completely overlapping with the second packet of the mobile grain-particle envelope. 

\begin{figure}[h]
     \includegraphics[width=0.45\textwidth]{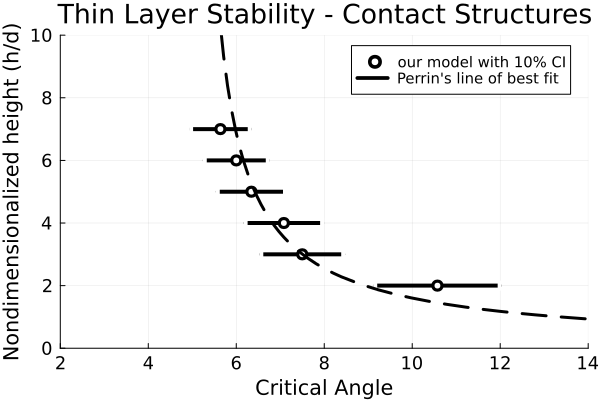}
     \caption{Critical angles for simulations of two to seven particles when using contact structures to pick temporal bounds for impulse calculation. }
     \label{fig:too_good}
 \end{figure}

While it is tempting to use these contact structures to inform temporal bounds for impulse calculation, it is difficult to tie these to physical intuition and make them consistent for the two particle case, the transition lengths, and the long particle chains. For example, using the first dimensionless time greater than one where the contact structures swing from the total chain length to one, produces a remarkable fit for the transition zone  (Figure \ref{fig:too_good}), however the physical intuition is murky, and it is difficult to reconcile with original two particle scenarios and the longer particle chains.

\begin{figure}[h]
     \includegraphics[width=0.45\textwidth]{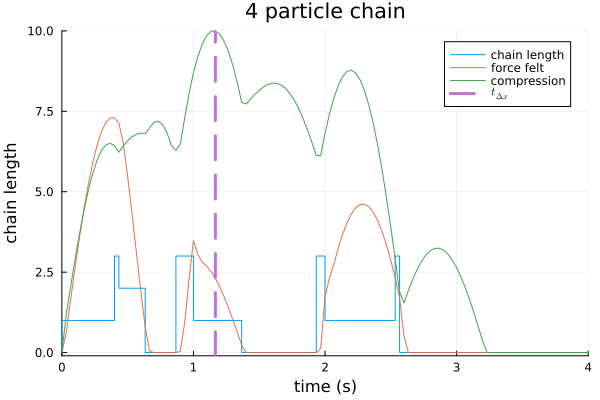}
     \caption{Visualization of contact structure (blue), forces felt (orange), and compression of the mobile grain for a 4 particle chain. The purple dashed line indicates where the particle is maximally compressed, which sets the timescale for our impulse calculation.}
     \label{fig:t_select}
 \end{figure}

A simpler choice for impulse timescale would be the periodicity for the mobile grain to collide and rebound. However, the idea of rebound becomes complicated once the  rebound effects become partitioned into increasingly separated mobile grain-particle and mobile grain-chain interactions. However, what remains clear is the initial compression, so we identify the time, within $3t_0$ at which the chain is maximally compressed, which represents half of the periodicity and then double this value. The bound of $3t_0$ is placed to ensure we are looking at compressions directly tied to the initial collision and not effects from compounding compression and gravity later in the simulation.  Figure \ref{fig:t_select} shows an example of this time being selected for a three particle simulation, and these are the temporal bounds used to produce our impulse calculations in figure \ref{fig:impulses}.  The fact that this approach does not directly reconcile with the contact structures, and the fact that the three particle simulation is a unique overlap of different contact dynamics, is our best guess as to why it fails to capture the three particle scenario.  We believe the fact that these tricksome transition lengths are of lengthscales typically associated with the granular mesoscale  is of no coincidence. Important future work would  combine information encoded in the contact structures and physical intuition to find a consistent definition of the impulse timescale which also sheds light on the evolution of dynamics within these transition lengths. 

\bibliography{apssamp}

\end{document}